\documentclass[sigplan]{acmart}

\acmConference[]{}
\acmYear{}
\acmISBN{} 
\acmDOI{} 
\startPage{1}

\setcopyright{none}
\copyrightyear{}           

\newcommand{\ignore}[1]{}
\usepackage{fancyhdr}
\usepackage{tabularx} 
\usepackage{amsfonts}
\usepackage[modulo]{lineno}
\usepackage[normalem]{ulem}
\usepackage{paralist}
\usepackage{todonotes}
\usepackage{booktabs}   
\usepackage{subcaption} 

\def\qed{\unskip\kern 10pt{\unitlength1pt\linethickness{.4pt}\framebox(6,6){}}}

\def\spha{\textsc{SPha}}

\def\powmon{\textsf{PowMon}}

\newtheorem{theorem}{Theorem}[section]
\newtheorem{definition}[theorem]{Definition}
\newtheorem{example}[theorem]{Example}

\begin{document}

\title[Compiler-assisted Adaptive Program Scheduling in big.LITTLE Systems]{Compiler-assisted Adaptive Program Scheduling in big.LITTLE Systems}         


\author{Marcelo Novaes}
\affiliation{
  \department{Department of Computer Science}              
  \institution{UFMG}            
  \country{Brazil}                    
}
\email{marcelonovaes@dcc.ufmg.br}          

\author{Vin\'{i}cius Petrucci}
\affiliation{
  \department{Department of Computer Science}              
  \institution{UFBA}            
  \country{Brazil}                    
}
\email{vinicius.petrucci@dcc.ufba.br}          

\author{Abdoulaye Gamati\'{e}}
\affiliation{
  \department{LIRMM}              
  \institution{CNRS}            
  \country{France}                    
}
\email{abdoulaye.gamatie@lirmm.fr}          

\author{Fernando Quint\~{a}o}
\affiliation{
  \department{Department of Computer Science}              
  \institution{UFMG}            
  \country{Brazil}                    
}
\email{fernando@dcc.ufmg.br}          

\begin{abstract}
Energy-aware architectures provide applications with a mix of low (LITTLE) and high (big)
frequency cores.
Choosing the best hardware configuration for a program running on such an architecture is
difficult, because program parts benefit differently from the same hardware configuration.
State-of-the-art techniques to solve this problem adapt the program's execution to dynamic
characteristics of the runtime environment, such as energy consumption and throughput.
We claim that these purely dynamic techniques can be
improved if they are aware of the program's syntactic structure.
To support this claim, we show how to use the compiler to partition source code
into program phases: regions whose syntactic characteristics
lead to similar runtime behavior.
We use reinforcement learning to map pairs formed by a program phase and
a hardware state to the configuration that best fit this setup.
To demonstrate the effectiveness of our ideas, we have implemented the Astro system.
Astro uses Q-learning to associate syntactic features of programs with hardware
configurations.
As a proof of concept, we provide evidence that Astro outperforms GTS, the ARM-based
Linux scheduler tailored for heterogeneous architectures, on the parallel benchmarks from
Rodinia and Parsec.
\end{abstract}



\maketitle

\section{Introduction}
\label{sec:int}

Contemporary hardware found in mobile phones and data centers sport multiple
ways to reduce energy consumption.
Two of these techniques are the combination of low and high power cores (the so
called big.LITTLE architectures)~\cite{Chung12}, and the ability to adjust power and speed
dynamically (DVFS)~\cite{LeSueur10}.
This design gives us the possibility to allocate to each parallel application
the hardware configuration that best suits it.
A hardware configuration consists of a number of cores, their type and their
frequency level.
We say that a configuration $H_1$ suits a program better than another 
configuration $H_2$ if $H_1$ runs said program more efficiently than $H_2$,
according to some metric such as runtime or energy consumption.
Nevertheless, even though we have today the possibility of choosing among several
configurations, the one that better fits the needs of a certain program, we still
have no clear technique to perform this choice seamlessly.

We call the task of allocating parts of a parallel program to processors the {\em code
placement problem}.
State-of-the-art approaches solve this problem dynamically or statically.
Dynamic solutions~\cite{Margiolas16,Nishtala17,Petrucci15} are implemented at
the runtime level, at the operating system, or via a middleware.
Static approaches~\cite{Grewe11,Mendonca17,Nugteren14,Verdoolaege13} are
implemented at the compiler level.
The main advantage of the dynamic approach is the fact that it can use runtime information to
weight the choices it makes.
Static techniques, in turn, provide reduced runtime cost and better leverage of
program characteristics.
In this paper, we claim that it is possible to join these two approaches,
achieving a synergy that, otherwise, could not be attained by each technique
individually.

To fundament this claim, we start from a technique that has been
proven effective to schedule computations in big. LITTLE architectures:
{\em Reinforcement learning}.
Nishtala {\em et al.}~\cite{Nishtala17} showed that reinforcement
learning helps to find good hardware configurations to applications subject
to varying dynamic conditions.
The beauty of this approach is adaptability: it provides the means to explore a vast universe
of states, formed by different hardware setups and runtime data changing over time.
Given enough time, well-tuned heuristics find a set 
of scheduling decisions that suits the underlying hardware.
Yet, ``enough time" can be too long.
The universe of runtime states is unbounded, and program
behavior is hard to predict without looking into its source code.
To speedup convergence, we resort to the compiler.

The compiler gives us two benefits.
First, it lets us mine program features, which we can use to train the learning
algorithm.
Second, it lets us instrument the program.
This instrumentation allows the program itself to provide feedback to the
scheduler, concerning the code region currently under execution.
Based on previous knowledge, collected statically, about characteristics of that
region, the scheduler can take immediate action.
An action consists in choosing a new state to represent program behavior,
and collecting the reward related to that choice.
Such feedback is then used to fine-tune and improve scheduling decisions.
As we show in Section~\ref{sec:eval}, convergence is faster, and runtime
shorter.

To validate our ideas, we have materialized them into a framework to
instrument and execute applications in heterogeneous architectures:
the {\em Astro System}.
Astro collects syntactic characteristics from programs and instruments them
using LLVM~\cite{Lattner04}.
Experiments in programs from Parsec~\cite{Bienia08} and Rodinia~\cite{Che09} running on an
Odroid XU4 show that we can obtain speedups of more than 10\% over the default GTS scheduler
used in ARM-based systems.
Such numbers result from the following contributions:
\begin{compactdesc}

\item[Observations:]
in Section~\ref{sec:mot}, we demonstrate that the performance of a program running on
a heterogeneous architecture vary depending on which part of its text we consider.
This observation points us to the key insight: the possibility of
augmenting an adaptive runtime apparatus with awareness of program
characteristics.

\item [Compiler:]
in Section~\ref{sub:partitioning}, we explain how to collect and discretize
program features, and in Section~\ref{sub:learning}, we explain how to
instrument a program, so to use said features to fine-tune an
adaptive code placement algorithm.

\item [Runtime:]
in Section~\ref{sub:run}, we show how to integrate the static
information that we collect with an adaptive runtime environment.
Once we train a program, we generate code that maps different parts of it to suitable hardware
configurations.
\end{compactdesc}


\section{Empirical Observations}
\label{sec:mot}

\begin{figure}[t!]
\begin{center}
\includegraphics[width=1\columnwidth]{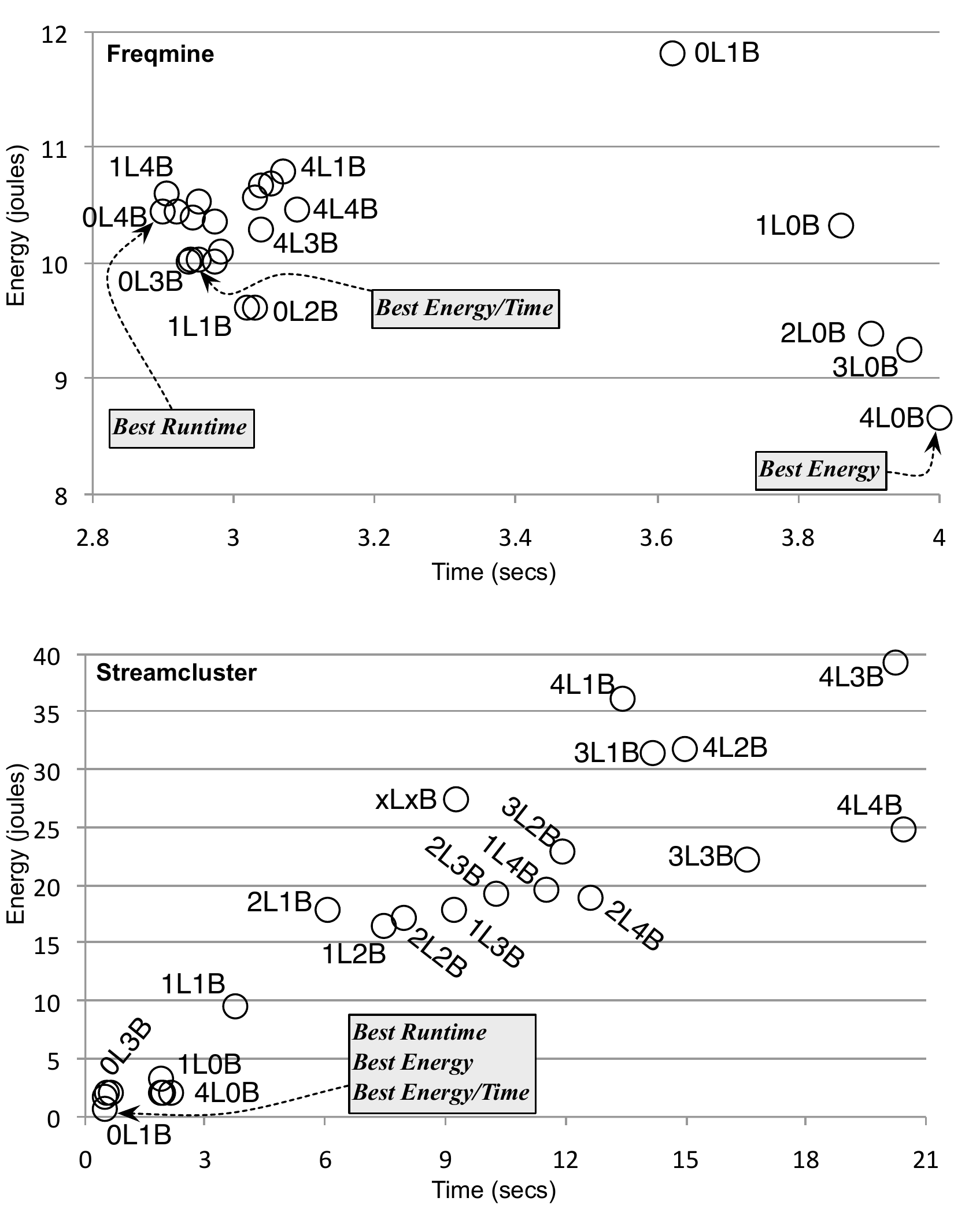}
\caption{Energy vs Processing time spent by two PARSEC benchmarks using
\textsf{simsmall} inputs. The notation \textsf{xLyB} denotes \textsf{x}
\underline{L}ITTLE cores, and \textsf{y} \underline{b}ig cores.}
\label{fig:cartesian_Parsec}
\end{center}
\end{figure}

\begin{figure}[t!]
\begin{center}
\includegraphics[width=1\columnwidth]{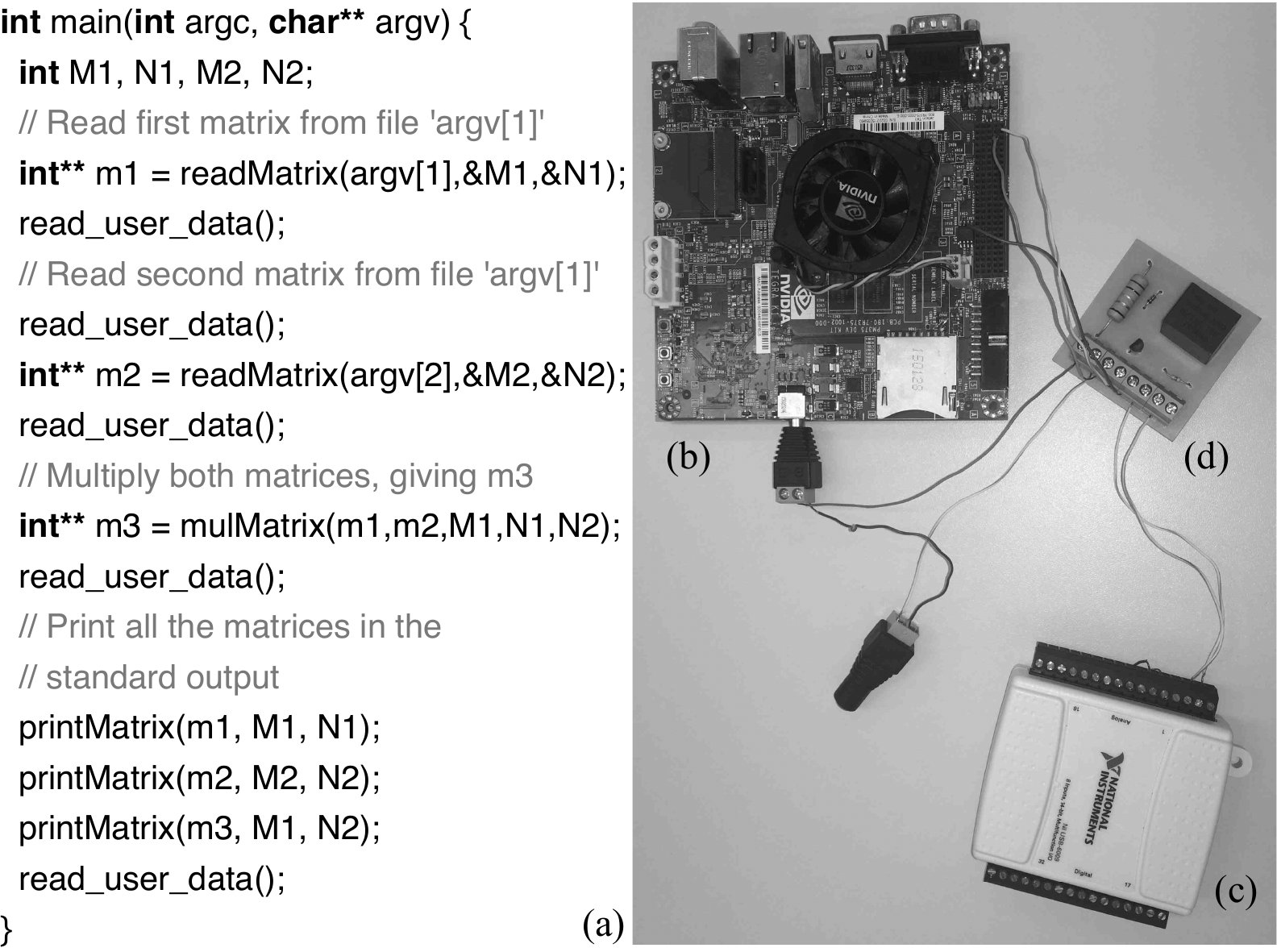}
\caption{
(a) Simple matrix multiplication implemented in C.
(b) The Nvidia TK1 board.
(c) NI 6009 Data Acquisition Device.
(d) Synchronization circuit.}
\label{fig:ex_prog_matMul}
\end{center}
\end{figure}

\begin{figure}[t!]
\begin{center}
\includegraphics[width=1\columnwidth]{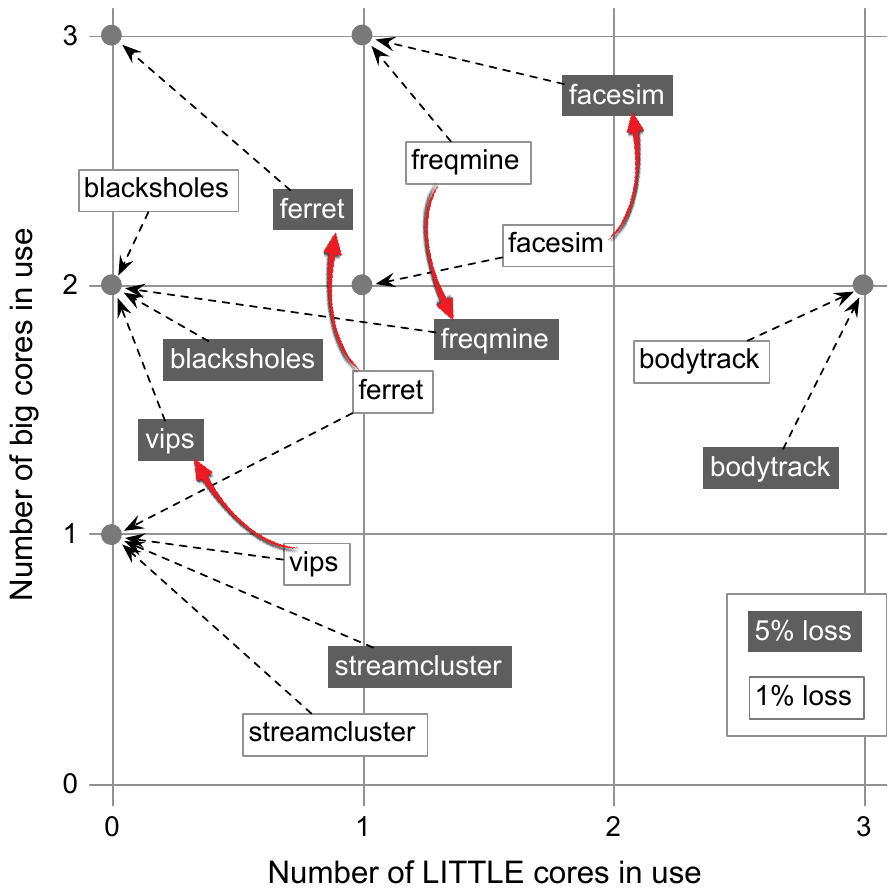}
\caption{Best configurations for seven PARSEC applications, if we accept an
slowdown of 1\% or 5\% to save more energy.}
\label{fig:best_Configs}
\end{center}
\end{figure}

This section motivates our work through three empirical observations.
First, different hardware configurations yield very different 
tradeoffs between power consumption and runtime speed for a program
(Figure~\ref{fig:cartesian_Parsec}).
Second, this behavior happens because programs 
have {\em power phases}: depending on the operations that they perform, they 
might consume more or less power per time unit (Figure~\ref{fig:ex_prog_matMul}).
Third, the best hardware configuration for a program might not 
suit the needs of a different application (Figure~\ref{fig:best_Configs}).
Central to the discussion in this section is the notion of a {\em hardware 
configuration}:

\begin{definition}[Hardware Configuration]
\label{def:config}
A heterogeneous architecture is formed by a set $P = \{p_1, p_2,$ $\ldots, p_n\}$
of $n$ processors. 
A hardware configuration is a function $H : P \mapsto \mathit{Boolean}$.
If $H(p_i) = \mathbf{True}$, then processor $p_i$ is said to be {\em active}
in $H$, otherwise it is said to be {\em inactive}.
\end{definition}

\noindent
\textbf{First Observation.} 
The same application might benefit differently from different hardware configurations.
This benefit is measured in terms of processing time and energy consumption.
Figure~\ref{fig:cartesian_Parsec} shows how two benchmarks from the PARSEC suite
-- \textsf{Freqmine} and \textsf{Streamcluster} -- fare on an Odroid XU4 board featuring 4
Cortex-A15 2.0Ghz cores and 4 Cortex-A7 1.4Ghz cores.
Following a nomenclature adopted by ARM, we shall call the A15 cores {\em bigs},
and the A7 cores {\em LITTLEs}.
By switching on and off the different cores, we have 24 different
hardware configurations\footnote{We have $24 = 5 \times 5 - 1$ configurations,
because we do not count the setup in which all cores are off.}

Each dot in the figure represents the average of 10 executions on the same
configuration, using the smallest\footnote{This experiment would take
12 days using the largest inputs.} input available in PARSEC.
Variance is almost negligible, staying under 1\% in every sample, for the
two benchmarks.
The X-axis shows the sum of the execution times of processors active in a
particular configuration; hence, it is not clock time.
Energy is measured with the Odroid XU3 
on-board power measurement circuit
and refers to work performed within the processors only; thus, peripherals are
not considered.

Figure~\ref{fig:cartesian_Parsec} lets us conclude that the energy and runtime
footprint of applications vary greatly across different hardware configurations.
For instance, the most time efficient configuration for \textsf{Freqmine} is
\textsf{0L4B}, i.e., four bigs and no LITTLEs (2.90secs, 10.43J).
However, the most energy efficient configuration is \textsf{4L0B} (4.01secs,
and 8.65J).
Results are not the same for \textsf{Streamcluster}.
The best energy configuration is \textsf{0L1B} (0.48secs, 0.69J).
This is also the most time efficient configuration.
\textsf{Freqmine} shows more parallelism than \textsf{Streamcluster}; therefore,
it benefits more from a larger number of cores.
This diversity of scenarios happen because programs have {\em phases}.
Energy and runtime behavior are similar within the same phase, and potentially
different across different phases.

\noindent
\textbf{Second Observation.} 
The instantaneous power consumed by a program is not always constant.
In other words, a program has {\em power phases}.
Figure~\ref{fig:ex_prog_matMul} (a) shows a program which we have crafted to emphasize the
different phases that a program undergoes during its execution.
This program performs the following actions:
(i) read two matrices from text files; (ii) multiply them and
(iv) prints all the matrices in the standard output.
In between each of these actions we have interposed commands to read data from
the standard input.

\begin{figure}[hbt]
\begin{center}
 \includegraphics[width=1\columnwidth]{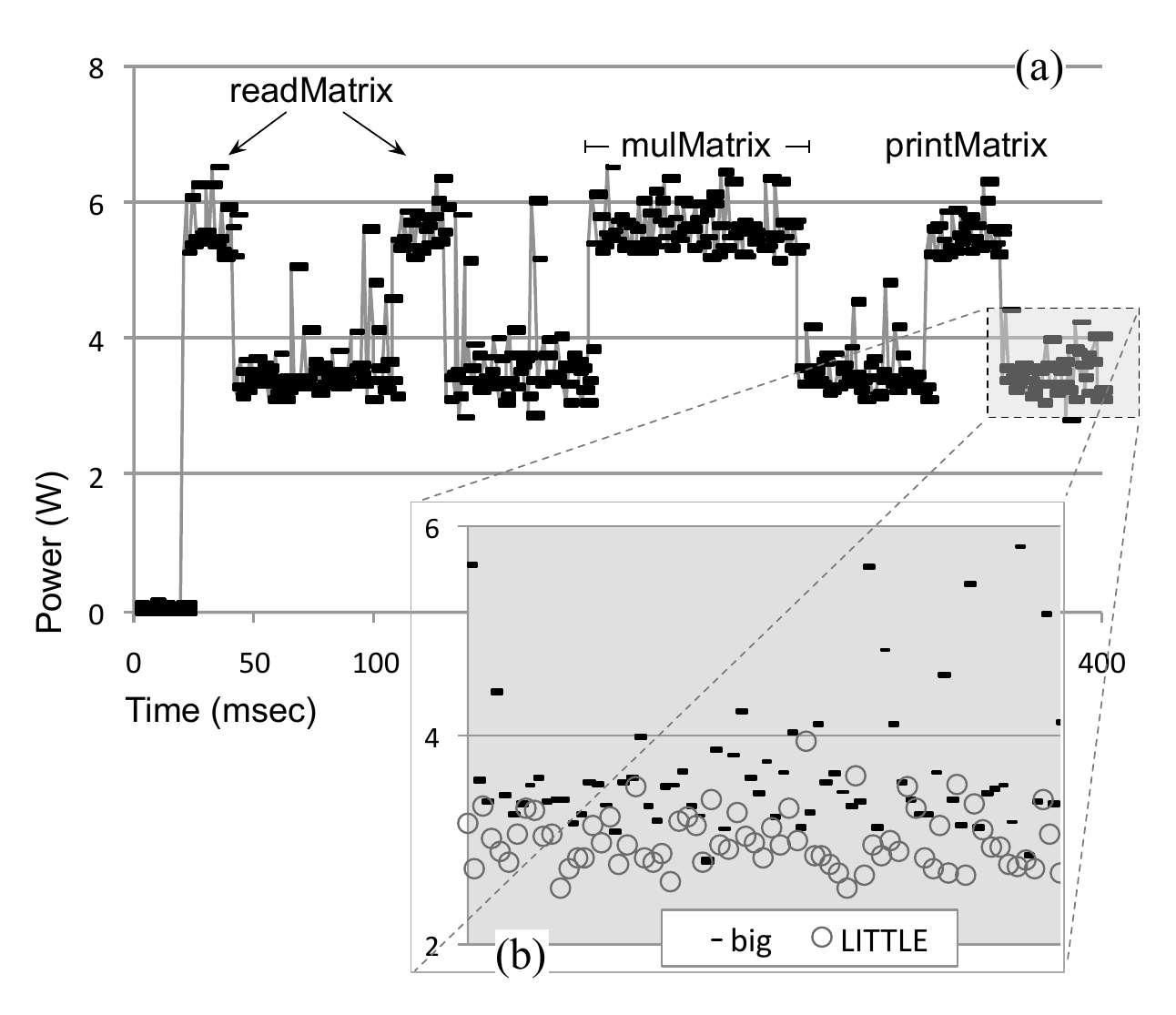}
\caption{
(a) Power profile of program seen in Figure~\ref{fig:ex_prog_matMul}.
The NI 6009 sample rate was 1000 samples/sec.
(b) Zoom of the power profile obtained during the last phase of the
program.}
\label{fig:matMul_Chart}
\end{center}
\end{figure}

Figure~\ref{fig:matMul_Chart} shows the power profile of this program.
This chart has been produced with JetsonLeap~\cite{Bessa16}, an apparatus that
let us measure the energy consumed by programs running on the Nvidia TK1 Jetson
board\footnote{In this section we use two different experimental setups: Odroid
XU4 and Tegra TK1. The former gives us the richness of configurations seen in
Figure~\ref{fig:cartesian_Parsec}. This diversity is absent on the latter, that
has only one LITTLE core. However, the TK1 board gives us access to JetsonLeap,
and, consequently, the ability to measure energy per programming events.}.
JetsonLeap is formed by three components: the target Nvidia board
(Figure~\ref{fig:ex_prog_matMul} (b)), a data acquisition device, which reads
the instantaneous power consumed by the board (Figure~\ref{fig:ex_prog_matMul} 
(c)), and a synchronization circuit, which lets us communicate to the power
meter which program event is running at each instant
(Figure~\ref{fig:ex_prog_matMul} (c)).

Distinct phases exist within the same program because it might use the hardware
resources differently, depending on which part of it is running.
By reading performance counters, we know that during matrix multiplication, CPU
is at is maximum usage.
During the input/output operations, this utilization drops slightly, and other 
components of the hardware, such as its serial port, are more exercised instead.
This fall is steep once the program is waiting for user inputs.
The CPU is not the only hardware component that accounts for power dissipation.
The JetsonLeap apparatus measure energy for the entire hardware.
Thus, the under utilization of the CPU does not mean that overall power 
consumption will decrease.
Nevertheless, variations in the CPU usage are likely to cause
variations in the power profile of the program.

Discovering such program phases by means of purely dynamic techniques is
possible, yet difficult.
As we shall demonstrate in Section~\ref{sec:eval}, we can use profiling
techniques, \`{a} la Hipster~\cite{Nishtala17}, to identify variations in
program behavior.
However, this approach has two shortcomings.
First, distinct program parts, with very different resource demands in
terms of memory, CPU, disk and such, can display similar dynamic characteristics.
For instance, we could imagine a scenario in which function
\textsf{read\_user\_data}, in Figure~\ref{fig:ex_prog_matMul} is implemented
via busy waiting.
In this case, instead of the valleys observed in Figure~\ref{fig:matMul_Chart},
we would encounter a power line similar to that produced by CPU-intensive
functions like \textsf{mulMatrix}.
Second, profiling-based techniques face a tradeoff between precision and
overhead.
Fast detection asks for high sampling rates; thus burdening the application
which originally we intended to optimize.
On the other hand, purely static approaches are not better either.
Although likely to yield lower adaptation overhead, they fail to account for
information only available at runtime such as varying input sizes.
For instance, a static scheduler might decide always run \textsf{mulMatrix} and
\textsf{read\_user\_data} in different configurations.
However, when operating on matrices that are too small, the cost of changing the
hardware configuration might already overshadow the possible gains available
through more parsimonious usage of the architecture's resources.

\noindent
\textbf{Third Observation.} 
The best architecture configuration, in terms of runtime or energy consumption, 
differs among programs.
Figure~\ref{fig:best_Configs} shows the best configurations that we have found
on the Odroid XU4 setup, for six PARSEC applications.
We define the best configuration as the one that spends
less energy, given a certain slowdown compared to the fastest configuration.
Clearly, there is not a single winner.
Configurations vary among programs, and even within the same program, given
different acceptable slowdowns.

In the rest of this paper, we shall describe a general methodology, henceforth
called the Astro system, which mixes static and dynamic analyses, to find good
hardware configurations for the functions invoked during the
execution of a program.
In this section, we have highlighted key motivation behind our design:
(i) a modern heterogeneous hardware exposes a number of different
configurations that is too large to be evaluated manually;
(ii) a program presents power phases, which can be more easily detected by
methods that are aware of structural properties of the code.
Thus, we claim that effective adaptation demands knowledge of program
characteristics.
Such information is readily available to the compiler; however, it is hard to be
precisely acquired by techniques unaware of the program's structure.

\section{The Astro System}
\label{sec:sol}

This section describes the design and implementation of our approach to
solve the problem of finding good hardware configurations for programs.
We state this problem as follows:

\begin{definition}
\label{def:spha}
\textsc{Scheduling of Programs in Heterogeneous Architectures} (\spha) \\
\textbf{Input}: a program $P$, its input $I$, hardware
configurations $H_1, \ldots H_n$, energy threshold $E$, and performance
threshold $S$.\\
\noindent
\textbf{Output}: $P'$, a new version of $P$, which switches between
configurations, and process $I$ using $E$\% less energy, with a slowdown of
no more than $S$\%.
\end{definition}

\begin{figure}[b!]
\begin{center}
\includegraphics[width=0.95\columnwidth]{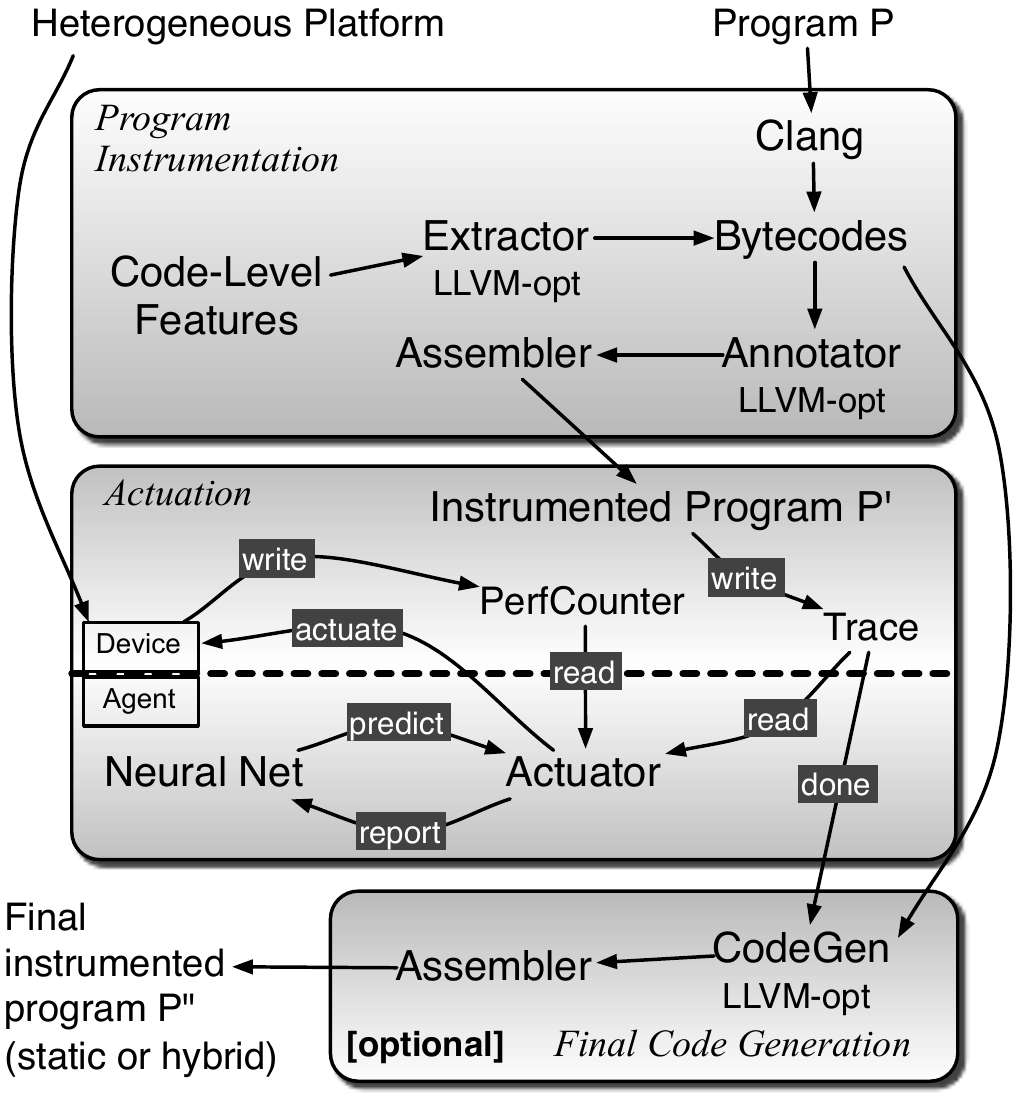}
\caption{The Astro System.}
\label{fig:General_Overview}
\end{center}
\end{figure}

In this paper, we solve \spha{} using
an assortment of techniques, which give us the means
to generate code that is well adapted to different architectures and
workloads.
Figure~\ref{fig:General_Overview} provides a general overview of these
techniques, emphasizing the different stages over which we go in the process of
solving \spha.
Section~\ref{sub:partitioning} describes program instrumentation, a necessary step to
partition a program into phases.
Section~\ref{sub:learning} goes over actuation; and
Section~\ref{sub:run} discusses the generation of the final program.
However, before we move into the particulars of our solution to \spha, we 
provide a brief introduction to Q-Learning, the flavour of reinforcement learning that
we have adopted.

\paragraph{Q-Learning.}
Q-learning is a reinforcement learning algorithm~\cite{Sutton:1998:IRL}.
Given some notion of state (Definition~\ref{def:state}) and reward
(Definition~\ref{def:reward}), it finds an optimized policy to perform the
best action (Definition~\ref{def:action}).
Q-learning is attractive because there is no need to know in advance the precise results of the actions before we perform them; that is, we learn about the environment as we perform actions on it.
A Markov Decision Process (MDP) drives Q-learning. A MDP is given by a set of states $S$, a set of possible actions $A$, a reward function
$R : S \times A \rightarrow \mathbb{R}$, and a state transition mapping $T: S \times A \rightarrow S$ that describes the effects of taking each action in each state of the environment.
The Markov property says that the results of an action depends only on the state
where the action was taken, regardless of any other prior states.

\subsection{Phase Partitioning}
\label{sub:partitioning}

A running program might cause the hardware to go over an infinite number of
different states.
Because this universe is unbounded, Definition~\ref{def:state} discretizes the
notion of a {\em State}.
In that definition, $S$ is a {\em Program Phase} and $D$ is a {\em Hardware
Phase}.
Program phases are discussed in Section~\ref{sss:st_phase}, and
hardware phases are discussed in Section~\ref{sss:dyn_phase}.

\begin{definition} [State]
\label{def:state}
A state is a triple $\langle H, S, D \rangle$ representing a
hardware configuration $H$, a program phase $S$ and a hardware phase $D$.
\end{definition}

\subsubsection{Program Phases}
\label{sss:st_phase}

Static Program Phases depend only on the syntax of a program.
Definition~\ref{def:prog_phase} formalizes this notion.
A static program phase is not equivalent to a {\em program
region},
because different regions can present the same set of feature ranges.
Example~\ref{ex:phases} clarifies the meaning of these definitions.

\begin{definition} [Program Phase]
\label{def:prog_phase}
A code-level feature (also called code feature or simply feature) is a syntactic
characteristic
of a program, such as number of n-nested loops or instruction mix.
A feature range is a contiguous interval of values that a feature can assume,
and that partitions the feature space into equivalence classes.
A program phase $S$ is a group of feature ranges, covering different
features.
\end{definition}

\begin{example}
\label{ex:phases}
The density of arithmetic and logical instructions is a code-level feature,
which we obtain by dividing the number of such opcodes by the total number of
program instructions.
We can define different feature ranges covering this metric, such as
$[0, 0.25)$, $[0.25, 0.50)$ and $[0.5, 1.00]$.
The number of nested loops yields another feature.
In this case, possible ranges are $[0, 1]$, $[2, 3]$ and $[4, +\infty]$.
Finally, an expectation on the number of I/O routines called in a function
gives us a third feature.
A heuristic to estimate it is: $\Sigma_i 10^n$, for
every I/O call $i$ nested into $n$ loops.
Potential intervals for this metric are $[0, 1)$, $[1, 10)$, $[10, 100)$ and $[100, +\infty]$.
The $3 \times 3 \times 4$ possible combinations of these ranges gives us 
36 program phases.
If we collect these features for each function in the program code, then we
can map any of them to one of these program phases.
\end{example}

In this paper, we mine (e.g., collect) features from the
intermediate program representation that the compiler manipulates before
producing executable code.
We have implemented a {\em Phase-Extractor} using the LLVM
compiler.
The result of mining program features is a map that assigns phases to
program regions.
This map depends on the choice of program region.
Many different granularities of regions are possible, such as instruction,
basic block, loop, Single-Entry-Single-Exit block~\cite{Ferrante87}, etc.
We have chosen to work mostly at the granularity of functions.
The ``mostly" in this case, refers to the fact that we also change phases
before and after library calls that cause the program to block waiting for
some event (see the \textsf{Barrier} phase, in the discussion that follows).
Pragmatically, this amounts to say that the instrumented program adds logic to
change phases at the entry point of functions, and around certain library
calls.

\begin{figure}[t!]
\begin{center}
\includegraphics[width=1\columnwidth]{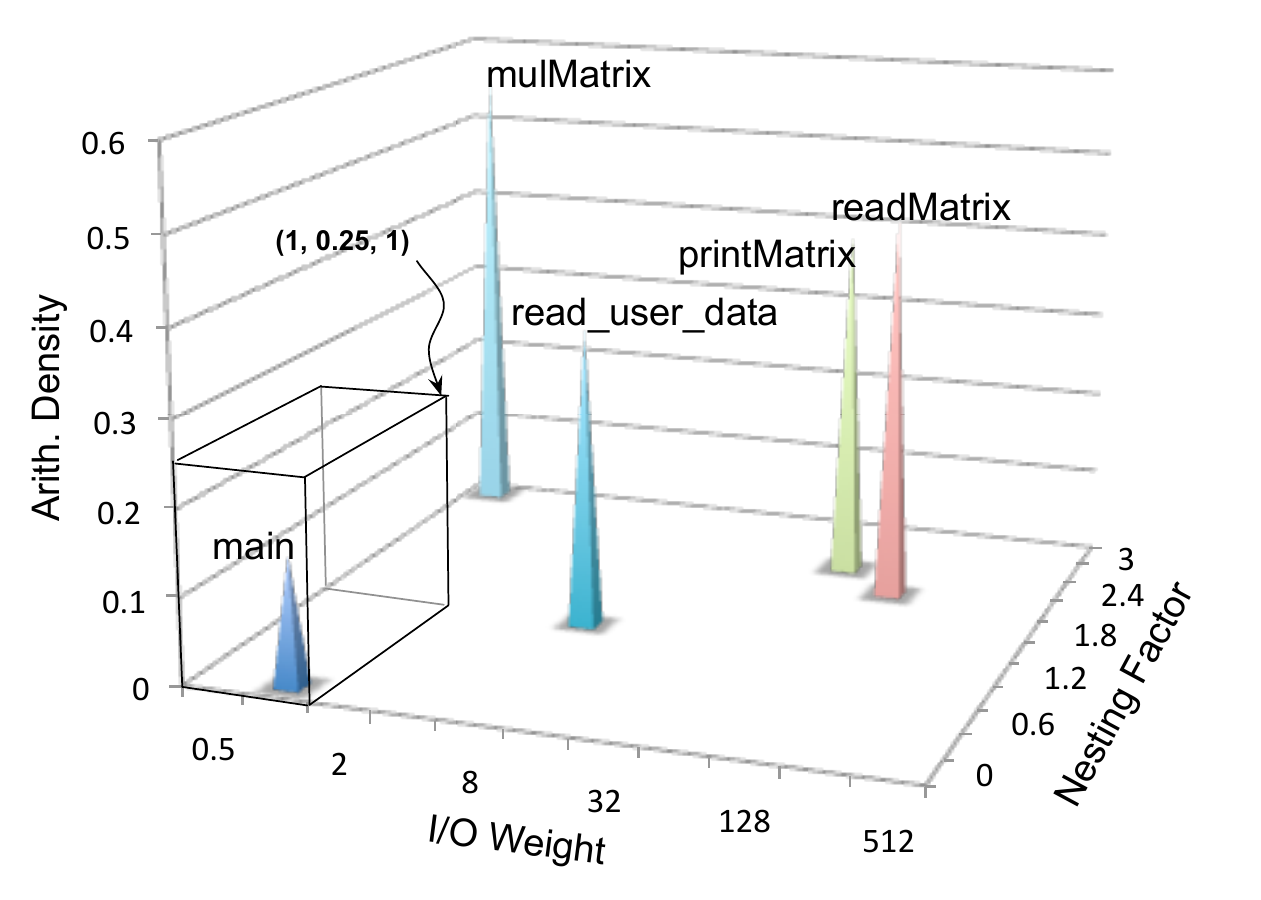}
\caption{Mapping the functions in Figure~\ref{fig:ex_prog_matMul} (a) to program
phases.}
\label{fig:ExPhasePartition}
\end{center}
\end{figure}

\begin{example}
\label{ex:regions}
Figure~\ref{fig:ExPhasePartition} shows the five functions in
Figure~\ref{fig:ex_prog_matMul}, classified according to features seen
in Example~\ref{ex:phases}.
We are assigning these functions hypothetical values.
Because we have three features, we can map them into a three-dimensional
space.
Each phase corresponds to a cube in this space.
Figure~\ref{fig:ExPhasePartition} shows the sub-space that corresponds to the
phase: $\mathsf{Arith. Density} \in [0, 0.25)$, 
$\mbox{\textsf{I/O Weight}} \in [0, 1)$ and
$\mathsf{Nesting Factor} \in [0, 1)$.
Function \textsf{main}, in our example, fits in this phase.
\end{example}

\noindent
\textbf{Our Choice of Program Phases.}
In our implementation, we combine four code features to determine program
phases.
These features are all ``densities", i.e., they represent a certain quantity of
instructions normalized by the total of instructions in the target function.
We use the following features:
\begin{compactitem}
\item \textsf{IO-Dens}: proportion of library calls that perform I/O 
operations;
\item \textsf{Mem-Dens}: proportion of instructions that access memory
(loads and stores);
\item \textsf{Int-Dens}: proportion of arithmetic and logic instructions
that operate on integer types.
\item \textsf{FP-Dens}: proportion of arithmetic and logic instructions
that operate on floating point types.
\item \textsf{Locks-Dens}: proportion of lock instructions.
\item \textsf{Barrier}: true when the program invokes a multi-thread barrier that forces it to
wait for some blocking event.
\item \textsf{Net}: true when the program invokes a library call that forces it
to wait for some network-related event.
\item \textsf{Sleep}: true when the program invokes a sleep library call that
forces it to wait unconditionally.

\end{compactitem}

We have defined four program phases, which appear as combinations
of the features above.
This choice is arbitrary.
We have opted for a simple partitioning, involving only a handful of
features for convenience, as this choice already lets us support the main
thesis of this paper: that static features greatly enhance the dynamic
scheduling of computations in heterogeneous hardware.
The program phases that we shall consider in Section~\ref{sec:eval} are:
\begin{compactitem}
\item \textsf{Blocked}: \textsf{Barrier} = true or \textsf{Net} = true or \textsf{Sleep}
= true or \textsf{Locks-Dens} $> 0.5$;
\item \textsf{I/O Bound}: \textsf{IO-Dens} + \textsf{Mem-Dens}
$> 0.5$ and not(\textsf{Blocked}) and \textsf{Locks-Dens} = 0;
\item \textsf{CPU Bound}: \textsf{Int-Dens} + \textsf{FP-Dens}
$> 0.5$ and not(\textsf{Blocked});
\item \textsf{Other}: in case none of the previous relations hold.
\end{compactitem}

\subsubsection{Hardware Phases}
\label{sss:dyn_phase}

While the program phases seen in Section~\ref{sss:st_phase} depend only on
syntactic program characteristics, {\em hardware phases} depend
on the dynamic state of the hardware:

\begin{definition}[Hardware Phase]
\label{def:hd_phase}
A Performance Counter is any monitor that collects dynamic information
about the hardware state, such as CPU performance and cache miss rate.
The domain over which the performance counter ranges can be partitioned into
phases.
Given a collection of performance counters $\{C_1,$ $C_2, \ldots, C_n\}$, where
each $C_i$ is partitioned into $R_i$ phases, then a hardware phase is any
combination within the product $R_1 \times R_2 \times \ldots \times R_n$.
\end{definition}

The monitoring of hardware phases does not require program instrumentation.
Instead, an {\em actuator} reads the state of hardware performance counters
periodically.
Modern architectures already provide an array of performance counters
that can be queried.
In this paper, we consider four kinds of counters to define hardware phases:
\begin{compactitem}
\item \textsf{IPC}: instructions per cycle in the ranges $[0, .5), [.5,$
$1.0), [1.0, +\infty)$;
\item \textsf{CMA}: cache misses per cache accesses in the ranges
$[0, 1\%), [1\%, 5\%), [5\%, +\infty)$;
\item \textsf{CMI}: cache misses per instruction executed, in the ranges
$[0, .1\%), [.1\%, .5\%), [.5\%, +\infty)$;
\item \textsf{CPU}: utilization of the CPU, in the ranges
$[0, 20\%),$ $[20\%, 50\%), [50\%, +\infty)$.
\end{compactitem}
Each counter is partitioned in three buckets.
Therefore, we consider a total of $3 \times 3 \times 3 \times 3 = 81$
hardware phases.

\subsection{Actuation}
\label{sub:learning}

\begin{figure}[t!]
\begin{center}
\includegraphics[width=1\columnwidth]{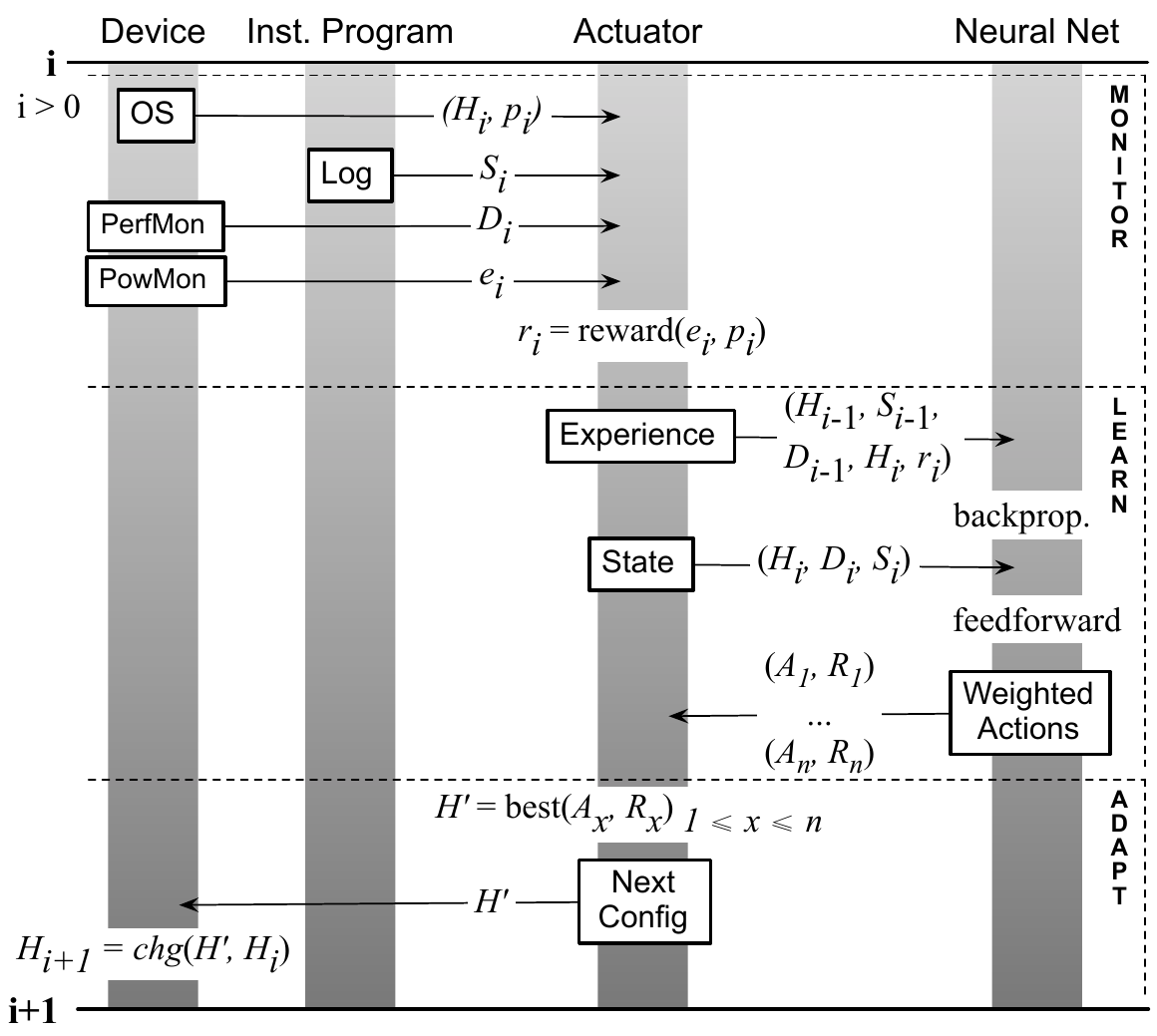}
\caption{The Actuation Algorithm.}
\label{fig:Algorithm}
\end{center}
\end{figure}

The heart of the Astro system is the Actuation Algorithm outlined in
Figure~\ref{fig:Algorithm}.
Actuation consists of {\em phase monitoring}, {\em learning} and
{\em adapting}.
These three steps happens at regular intervals, called {\em check points}, which,
in Figure~\ref{fig:Algorithm}, we denote by \textsf{i} and \textsf{i+1}.
The rest of this section describes these events.

\subsubsection{Monitoring}
\label{sss:monitoring}

To collect information that will be later used to solve \textsc{SPha}, Astro
reads four kinds of data.
Figure~\ref{fig:Algorithm} highlights this data:
\begin{compactitem}
\item From the Operating System (\textsf{OS}): current hardware configuration
$H$ and instructions $p$ executed since last check point.
\item From the Program (\textsf{Log}): the current program phase $S$.
\item From the device's performance counters (\textsf{PerfMon}): the current
hardware phase $D$.
\item From the power monitor (\powmon~\cite{Walker17}): the energy $e$
consumed since the last checkpoint.
\end{compactitem}
The monitor collects this data at periodic intervals, whose granularity is
configurable.
Currently, it is 500 milliseconds.
The recording of the program phase is aperiodic, following from instrumentation inserted in
the program by the compiler.
As discussed in Section~\ref{sss:st_phase}, information is logged at
the entry point of functions, and around library calls that might cause the program to enter a
dormant state.
The hardware configuration is updated whenever it changes.
The metrics $e$ and $p$ lets us define the notion of {\em reward} as follows:

\begin{definition}[Reward]
\label{def:reward}
The reward is the set of observable events that determine how well the learning
algorithm is adapting to the environment.
The reward is computed from a pair $(e, p)$, formed by the {\em Energy
Consumption Level} $e$, measured in Joules per second (Watt), and the {\em CPU
Performance Level} $p$, measured in number of instructions executed per second.
\end{definition}

The metric used in the reward is given by a weighted form of performance per watt, namely 
$MIPS^\gamma / Watt$, where $\gamma$ is a design parameter that gives a boosting performance 
effect in the system.
This is usually a trade-off between the performance and energy consumption.
To optimize for energy, we let $\gamma = 1.0$.
A value of $\gamma = 2.0$ emphasizes performance gains: the reward
function optimizes (in fact, maximizes the inverse of) the energy delay product per
instruction, given by $Watt / IPS^2$; letting $IPS = I/S$ we have
$(Watt \times S \times S) / I^2 = (Energy \times Delay) / I^2$.
This aims to minimize both the energy and the amount of time required to execute thread 
instructions~\cite{Brooks2000}.

\begin{figure}[t!]
\begin{center}
\includegraphics[width=1\columnwidth]{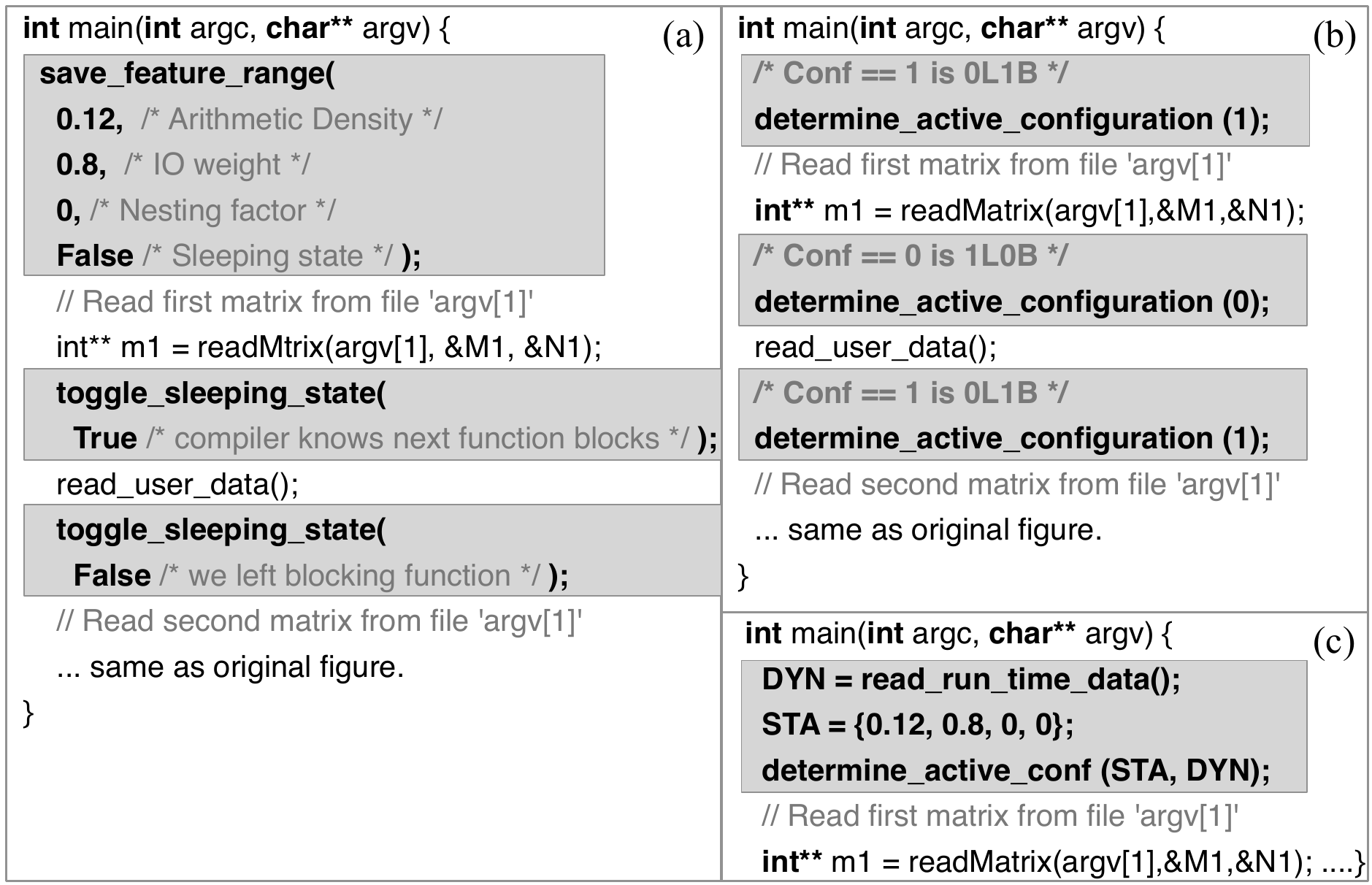}
\caption{
(a) Instrumentation to mine features.
(b) Final instrumentation, inserted in production code.}
\label{fig:Example_instrumentation}
\end{center}
\end{figure}

\begin{example}
Continuing with Example~\ref{ex:regions},
Figure~\ref{fig:Example_instrumentation} (a) shows the instrumentation of
function \textsf{main} (Figure~\ref{fig:ex_prog_matMul}) to log program phases.
\end{example}

\subsubsection{Learning}
\label{sss:learning}


The learning phase uses the Q-learning algorithm.
As illustrated in Figure \ref{fig:Algorithm}, a key component in this process is a multi-layer Neural Network (NN) that receives inputs collected by the
Monitor.
The NN outputs the actions and their respective rewards to the Actuator  so that a new system adaptation can be carried out.
Following common methodology, learning happens in two phases: {\em back-propagation} and
{\em feed-forwarding}.
During back-propagation we update the NN using the experience data given by the Actuator
(Figure ~\ref{fig:Algorithm}).
Experience data is a triple: the current state, the action performed and the reward thus
obtained.
The state consists of a hardware configuration $(H_{i-1})$, static features $(S_{i-1})$ and dynamic features $(D_{i-1})$ at check points \textsf{i-1}. The action performed at check point \textsf{i-1} makes the system move from hardware configuration $H_{i-1}$ to $H_i$. The reward is given by $r_{i}$, received after the action is taken.
The NN consists of a number of layers including computational nodes, i.e., neurons. The input layer uses one neuron to characterize each triple $(state,action,reward)$. The output layer has one neuron per action/configuration available in the system.
During the feed-forward phase, we perform predictions using the trained NN.
Each node of the NN is responsible to accumulate the product of its associated weights and
inputs.
Given as input a state $(H_{i}, D_{i}, S_{i})$ at check point \textsf{i}, the result of the feed-forward step is an array of pairs $A \times R$, where
$A$ is an {\em action}, and $R$ is its {\em reward}, estimated
by NN.
Actions determine configuration changes; rewards determine the expected
performance gain, in terms of energy and time, that we expect to obtain
with the change.
We use the method of gradient descent to minimize a loss function given by the difference
between the reward predicted by the NN, and the actual value found via hardware performance
counters.

\subsubsection{Adaptating}
\label{sss:adaptation}

At this phase, Astro takes an {\em action}.
Together with states and rewards, actions are one of the three core
notions in Q-learning, which we define below:

\begin{definition} [Action]
\label{def:action}
Action is the act of choosing the next hardware configuration $H$ to be
adopted at a given checkpoint.
\end{definition}

An action may change the current hardware configuration; hence,
adapting the program according to the knowledge inferred by the Neural Network.
Following Figure~\ref{fig:Algorithm}, we start this step by choosing, among
the pairs $\{(A_1, R_1), \ldots, (A_n, R_n)\}$, the action $A_x$ associated 
with the maximal reward $R_x$.
$A_x$ determines, uniquely, a hardware configuration $H'$.
Once $H'$ is chosen, we proceed to adopt it.
However, the adoption of a configuration is contingent on said configuration
being available.
Cores might not be available because they are running higher privilege jobs, 
for instance.
If the Next Configuration is accessible, Astro enables it; otherwise, the
whole system remains in the configuration $H_i$ active at check point \textsf{i}.
Such choice is represented, in Figure~\ref{fig:Algorithm}, by the function
$H_{i+1} = \mathit{chg}(H', H_i)$.
Regardless of this outcome, we move on to the next check point,
and to a new actuation round.


\subsection{Code Scheduling}
\label{sub:run}

After we have trained a program to a given architecture,
we imprint this knowledge directly in that program's code.
In Figure~\ref{fig:General_Overview}, this step is named {\em Final Code
Generation}.
Code generation consists in inserting instrumentation into the target program.
Instrumentation is inserted in the same regions modified to mark 
program phases (see Section~\ref{sss:st_phase}): at the entry point of
functions, and around particular library calls.
Example~\ref{ex:final_inst} illustrates this instrumentation.

\begin{example}
\label{ex:final_inst}
Figure~\ref{fig:Example_instrumentation} shows the final actuation code for the
program in Figure~\ref{fig:ex_prog_matMul}.
Function \textsf{determine\_active\_configuration} tries to move the program
to the configuration that has produced the largest rewards
for that program phase.
We consider two versions of instrumentation: static, as in
Figure~\ref{fig:Example_instrumentation}(b), and
hybrid, as in Figure~\ref{fig:Example_instrumentation} (c).
The latter can read hardware status to improve the decision making process.
\end{example}

The static scheduling discussed in Example~\ref{ex:final_inst} always maps the
same program region to the same hardware configuration.
Hybrid scheduling might change decisions, given enough runtime information.
As we show in Section~\ref{sec:eval}, the static scheduling yields lower
runtime overhead than Astro's hybrid scheduling.
However, this modus operandi is unable to adapt the program to its workload; and cannot
recover from bad decisions.
A striking example is the benchmark \textsf{ParticleFilter}
(see Fig.~\ref{fig:ExecutionResults} in Section~\ref{sub:actual}).
In this case, even with the runtime overhead,
the flexibility of hybrid instrumentation paid off in terms of energy and speed.

\section{Evaluation}
\label{sec:eval}

This section presents an experimental evaluation of the Astro system over
several parallel benchmarks running on a big.LITTLE system.
In the process of evaluating Astro, we shall provide answers to the following
research questions:

\begin{compactitem}
\item \textbf{RQ1}: How close can Astro be from an optimal oracle?
\item \textbf{RQ2}: How does Astro compare against fixed and immutable
best configuration choices?
\item \textbf{RQ3}: How does Astro compare against state-of-the-art schedulers?
\item \textbf{RQ4}: How does Astro behave on an actual device?
\item \textbf{RQ5}: How much does Astro increase code size?
\end{compactitem}

\noindent
\textbf{Experimental Setup.}
We use two experimental setups: {\em program traces}, henceforth called {\em simulation};
and an actual device, the Odroid XU4.
Experiments in Section~\ref{sub:sim} use simulation because they involve testing
exhaustively every hardware configuration.
Experiments in Section~\ref{sub:actual} run on an actual device: the Odroid XU4
development board with a big.LITTLE ARM processor (Samsung Exynos 5422) featuring 4 big cores
(Cortex-A15 2.0 Ghz) and 4 LITTLE cores (Cortex-A7 1.4 Ghz), running on Linux odroid 3.10.63,
using the ``performance" frequency governor, with cores at maximum speed.
This device was also used to produce the simulation traces.
We report CPU power consumption via \powmon~\cite{Walker17}.
Astro is implemented on LLVM 3.8.

\noindent
\textbf{Benchmarks.}
The simulation traces used in Section~\ref{sub:sim} were produced on Parsec's
\textsf{FluidAnimate}~\cite{Bienia08}.
Experiments on Section~\ref{sub:actual} use eight benchmarks from Rodinia and Parsec.
These are the only programs that we can currently instrument, as our LLVM module does not
recognize mangled C++ routines yet (to discover program phases such as I/O density --
Sec.~\ref{sss:st_phase}).
We used \textsf{FluidAnimate} to obtain the initial learning parameters; hence, we do not use
it for validation.


\subsection{Results in the Simulated Environment}
\label{sub:sim}

\begin{figure}[htb]
\begin{center}
\includegraphics[width=1\columnwidth]{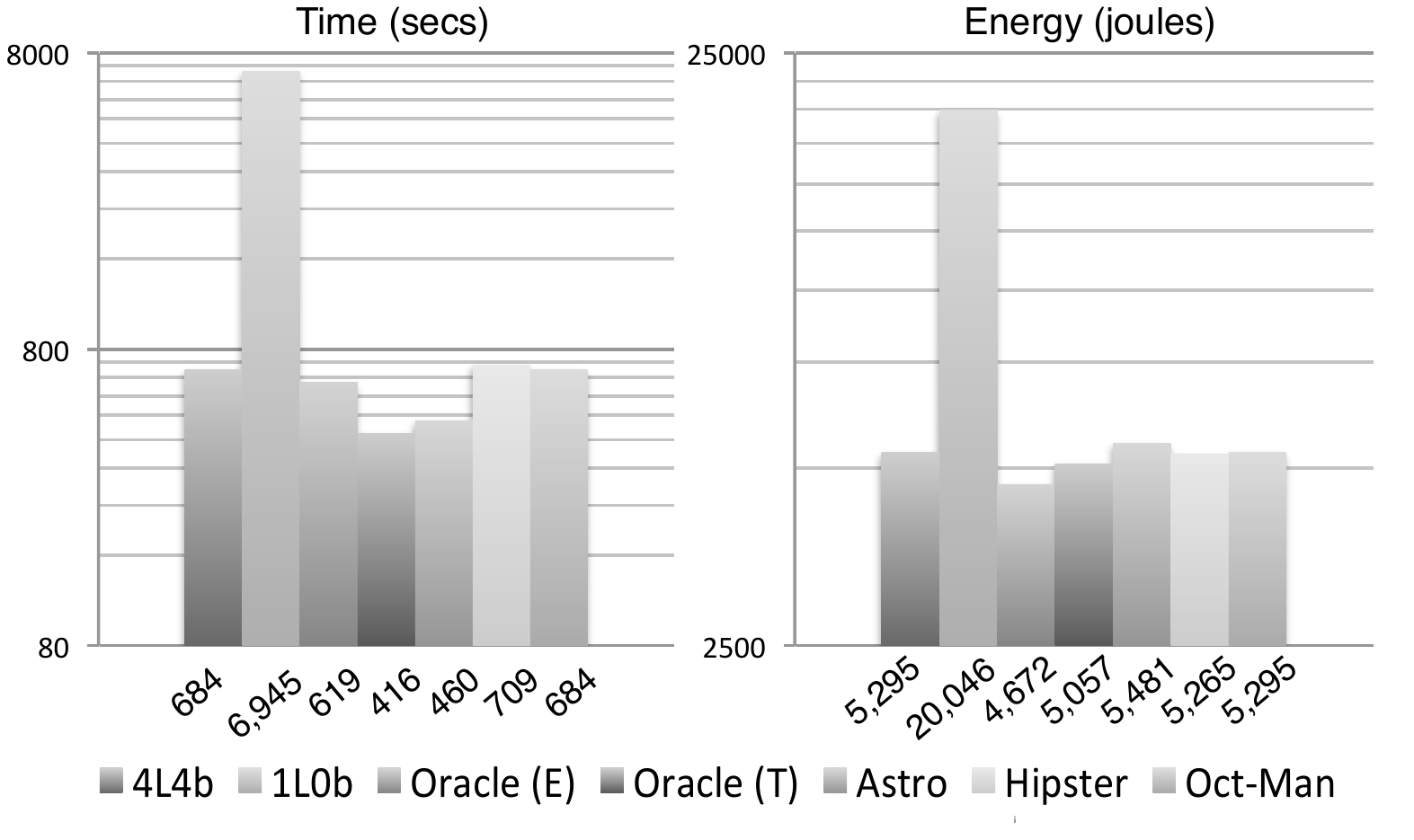}
\caption{Comparison between Astro and a system that chooses the next
configuration randomly.}
\label{fig:GeneralComparison}
\end{center}
\end{figure}

In this section we report results that are hard to obtain on an actual device, because they
involve exhaustive search on the universe of valid hardware configurations.
We have approximated the exhaustive execution of configurations by generating traces for every
hardware configuration.
These traces lets us simulate different behaviors, by choosing, at each
checkpoint, the reward offered by one of them.
Different policies can guide this choice:
optimal, best fixed and random for instance.
Producing such traces is time consuming, thus, we have produced them
only for \textsf{fluidanimate}.
We took between 410 seconds to up to 7,000 seconds to produce each trace, depending on the
hardware configuration.
Figure~\ref{fig:GeneralComparison} compares seven different scheduling
strategies built on top of this simulator, applied on \textsf{fluidanimate}.

\noindent
\textbf{RQ1: how close is Astro to an optimal oracle?} 
The data collected for every possible configurations lets us know, for each 
part
of the program, which configuration consumes less energy and has the best
performance.
We then combine these 24 traces into a single trace, choosing, at each
check point, a particular configuration.
This ``optimal" trace is what we call the {\em Oracle}.
Our oracle is not an optimal global solution to \spha.
Rather, it is a greedy approximation: given that at check-point $i$ we are
at configuration $H_i$, what is the configuration that gives us the best
reward at check-point $i + 1$.
Figure~\ref{fig:GeneralComparison} shows two oracles:
\textsf{(E)} and \textsf{(T)}.
The former yields optimal energy consumption; the latter yields optimal
execution time.
Astro's reward function prioritizes time over energy; hence, it leads to
execution times close to \textsf{T}.
If we schedule \textsf{Fluidanimate} with Astro, its final runtime is only
10\% slower than \textsf{T}.
However, it is more energy hungry: it uses 8\% more energy than \textsf{T},
and 15\% more energy than \textsf{E}.

\noindent
\textbf{RQ2: How does Astro compare against immutable
best configuration choices?} 
If we fix the hardware configuration, then \textsf{4b4L} (4 bit, 4 LITTLE
cores) gives us the best runtime and the best energy consumption for the
simulation of \textsf{Fluidanimate}.
This configuration is 45\% slower than Astro, yet it is 4\% more energy
efficient.
The fact that Astro, and the energy oracle, could beat \textsf{4b4L} is
surprising.
We have found out that 4b4L tends to slowdown programs at critical sections, due to an
excess of conflicts between threads.
Astro eventually learns to use configurations with less cores at these program phases;
hence, speeding up execution.
Figure~\ref{fig:GeneralComparison} also shows the
configuration that yields the slowest and more power hungry execution:
\textsf{1b0L}.
It is almost 15 times slower than Astro, and spends 3.6x more energy.

\noindent
\textbf{RQ3: How does Astro perform when compared with state-of-the-art
program schedulers?}
We tried to implement, on the simulator, two well-known schedulers for big. LITTLE
architectures: Hipster~\cite{Nishtala17} and Octopus-Man~\cite{Petrucci15}.
The implementation of Hipster used in Figure~\ref{fig:GeneralComparison} differs
slightly from the original description of Nishtala {\em et al}, although
we have reused much of their code base.
Hipster was originally conceived to deal with cloud workloads; hence, we had to
customize its state and reward function for multithreaded programs.
In this experiment, both, Hipster and Astro use the same reward function.
Octopus-Man is the profiling mechanism used in Hipster; hence, it does not
use the notion of reward.
Astro produces code that runs 17\% faster than Hipster, and 15\% faster than
Octopus-Man.
However, Astro uses 6\% more energy than the former, and 4\% more than the latter.

\subsection{Results in an Actual Device}
\label{sub:actual}

\begin{figure*}[htb]
\begin{center}
\includegraphics[width=1\textwidth]{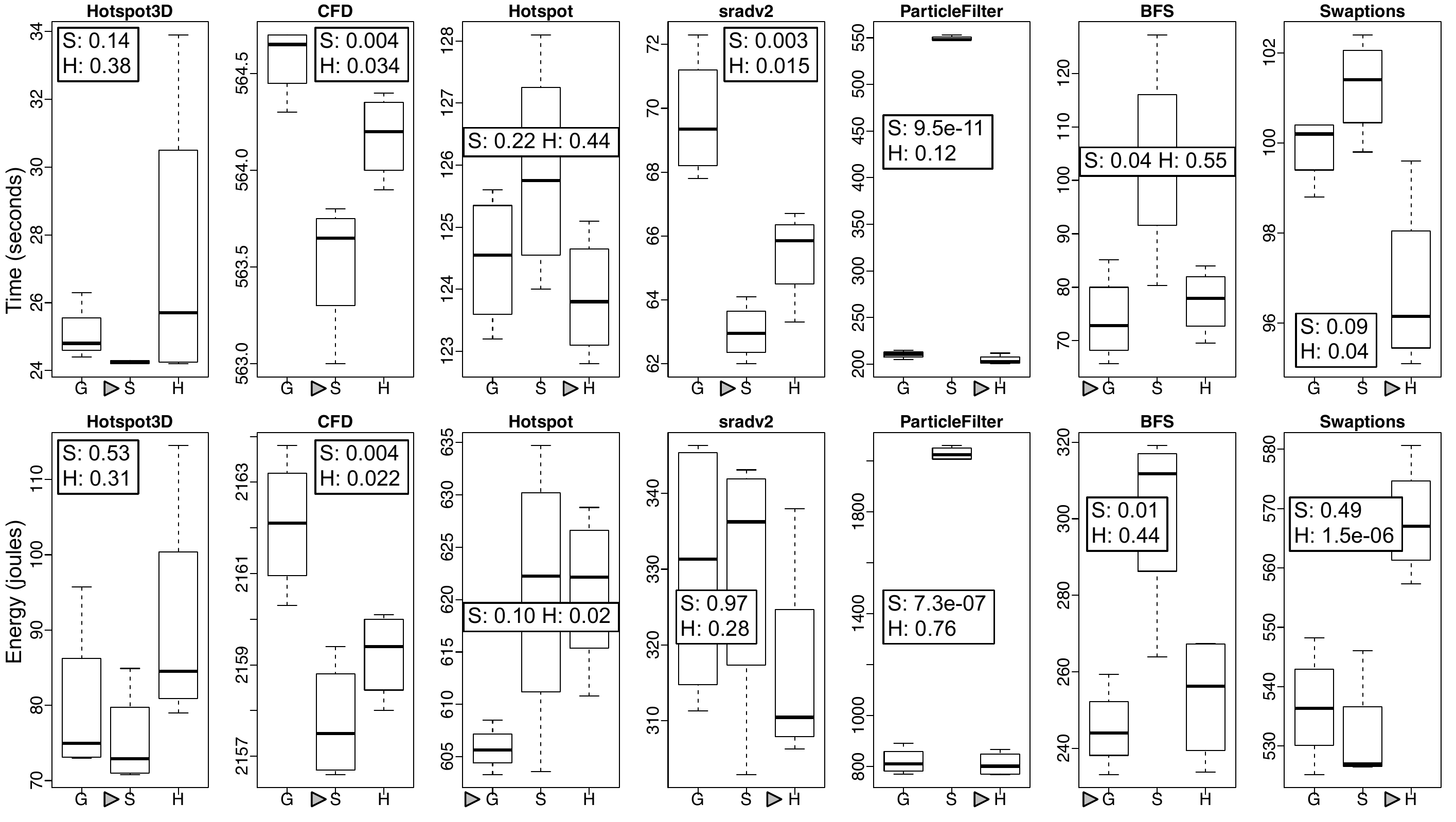}
\caption{Time (Top) and Energy (Bottom) comparison between
Astro and GTS (G).
``Static (S)" is the purely static version of Astro
(Fig.~\ref{fig:Example_instrumentation}b).
``Hybrid (H)" is the version that uses runtime information to improve on the static
decisions (Fig.~\ref{fig:Example_instrumentation}c).
Numbers in boxes are p-values for the \underline{S}tatic and \underline{H}ybrid
approaches, compared to \underline{G}TS.
Grey triangles indicate winning strategies.}
\label{fig:ExecutionResults}
\end{center}
\end{figure*}

\noindent
\textbf{RQ4: How does Astro behave on an actual device?}
Figure~\ref{fig:ExecutionResults} shows the runtime (5 samples) of three different
solutions to \spha: Astro (purely static or hybrid), and Global Task Scheduling (GTS).
GTS is a scheduling algorithm developed by ARM.
This scheduler is aware of the different
compute capabilities of big and LITTLE cores in the system.
It uses historical data of the running tasks and active cores to determine
where each individual thread will run. By tracking the load
information at runtime, GTS migrates tasks that are compute-intensive
to big cores and those that are less intensive to little cores. Load
balancing heuristics are periodically executed to minimize
concentrating compute-intensive threads excessively on big cores and
letting little cores underutilized.
Numbers reported for Astro include all the overhead of monitoring and
adapting the target application.

Astro, in its static or hybrid flavours, yields faster code than GTS in six benchmarks,
and more energy efficient code in five.
We show two p-values next to each plot: S and H.
The former is the probability that the static and purely dynamic (GTS) samples come from 
the same distribution.
The latter relates the hybrid and purely dynamic distributions.
The closer to zero, the more statistically significant are our results.
We emphasize that GTS is a state-of-the-art approach, widely used in
operating systems running on ARM hardware, and the fact that Astro can
consistently outperform it testifies in favour of the benefits of syntax awareness
when taking scheduling decisions.
There is no clear winner between the hybrid and static versions of Astro.
We observer that the former tends to be better in more regular (kernel-like) applications,
such as \textsf{CFD} and  \textsf{sradv2}.
We also observe strong correlation between runtime and energy consumption, except for
\textsf{Swaptions}.
In that case, the Static version of Astro tends to avoid using the high-frequency cores, a
fact that leads to slower runtime, but also to less power dissipation.
In \textsf{ParticleFilter} the static version was penalized for a wrong scheduling
decision: it stays in \textsf{1b2L}, and the lack of runtime information prevents it from
fixing this choice.

\begin{figure}[htb]
\begin{center}
\includegraphics[width=1\columnwidth]{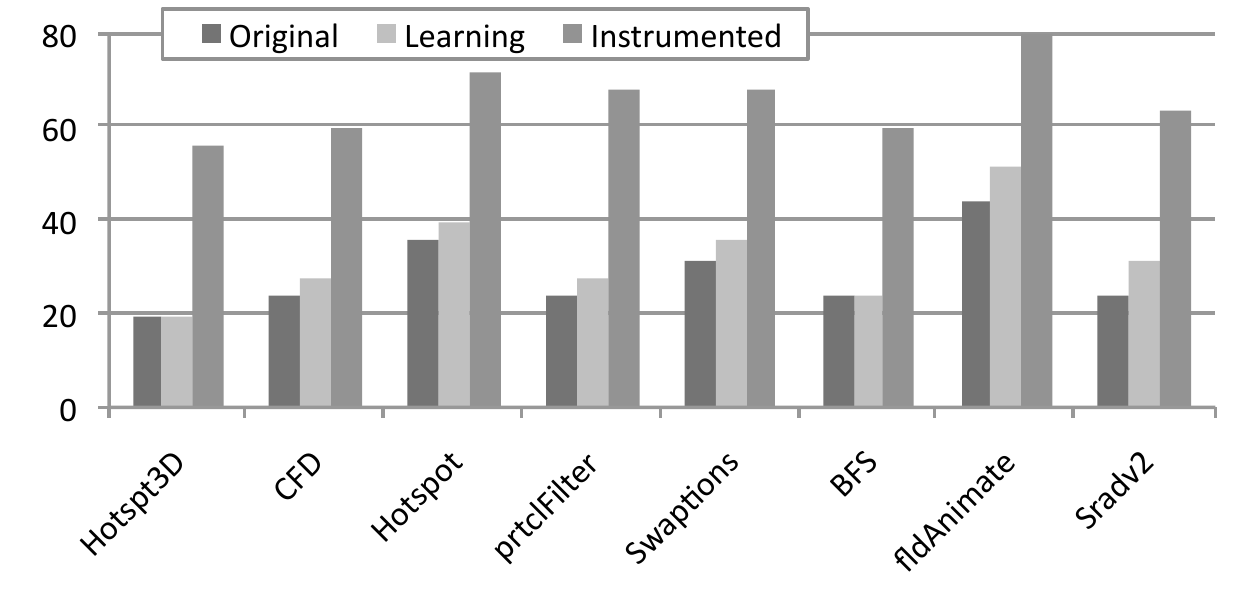}
\caption{Code size increase.
Y-axis shows code size (Kb).}
\label{fig:CodeSize}
\end{center}
\end{figure}

\noindent
\textbf{RQ5: How much does Astro increase code size?}
There are three different versions of instrumented programs:
those used during Astro's learning phase;
the programs that use static instrumentation; and
the programs that use hybrid instrumentation.
The binary size of the last two is the almost the same:
it consists of code that collects data, plus the Astro library.
The only different between static and dynamic instrumentation is the code used to collect
dynamic data in the latter version.
This different is too small; hence, in Figure~\ref{fig:CodeSize} we include both types of
binaries in the same bar: \textsf{Instrumented}.
As the figure shows, most of the size overhead imposed by Astro is due to its dynamic library.
This increase is constant across benchmarks.
The amount of instrumentation in binaries grows linearly with the program size.
This growth tends to be very small.
As evidence to this small growth, in the \textsf{Learning} phase, binaries do not use any
dynamically linked library; thus, code size expansion is due to instrumentation only, and it
is small, as seen in Figure~\ref{fig:CodeSize}.

\section{Related Work}
\label{sec:rw}

The problem of scheduling computations in heterogeneous architectures
(Definition~\ref{def:spha}) has attracted much attention in recent years.
Table~\ref{tab:relatedwork} provides a taxonomy of previous solutions to this
problem.
We group them according to how they answer each of the following four questions:
\begin{compactitem}
\item \textbf{Source}: is the program's code modified?
\item \textbf{Auto}: is user intervention required?
\item \textbf{Runtime}: is runtime information exploited?
\item \textbf{Learn}: is there any adaptation to runtime conditions?
\end{compactitem}
Perhaps the most important difference among the several strategies proposed to
solve \spha{} concerns the moment when they are used:
at compilation time, at runtime, or both.

\begin{table}[b]
\begin{footnotesize}
\centering
\begin{tabular*}{\columnwidth}{ @{\extracolsep{\fill}} rccccc}
{\em Work} & {\em Level} & {\em Source} & {\em Auto} & {\em Runtime} & {\em Learn} \\ \midrule
\cite{Poesia17} & C & Yes & Yes & No & Yes \\
\cite{Barik16}  & C & Yes & Yes & Yes & No \\
\cite{Rossbach13} & C/L & Yes & No & Yes & No \\
\cite{Luk09} & C/L & Yes & No & Yes & No \\
\cite{Joao12} & A/L & Yes & No & No & No \\
\cite{Lukefahr2016} & A & No & Yes & No & No \\
\cite{Craeynest12} & A & No & Yes & No & No \\
\cite{Nishtala17} & O & No & Yes & Yes & Yes \\
\cite{Petrucci15} & O & No & Yes & Yes & No \\
\cite{Augonnet11} & L & Yes & No & No & No \\
\cite{Piccoli14} & O/C & Yes & Yes & Yes & No \\
\cite{Tang2013} & O/C & Yes & Yes & Yes & No \\
\cite{Cong2012} & O/C & Yes & Yes & Yes & No \\
Astro & O/C & Yes & Yes & Yes & Yes \\ \bottomrule
\end{tabular*}
\caption{\label{tab:relatedwork}
Comparison between different solutions to \spha.
{\em Level}: at which level the technique is implemented: Architecture (A),
Operating System (O), Compiler (C) or Library/Programming model (L).
{\em Code}: ``Yes" if approach requires source code.
{\em Auto}: ``Yes" if it is performed automatically, without user intervention/annotation.
{\em Runtime}: ``Yes" if technique considers runtime information.
{\em Learn}: ``Yes" if technique adapts/learns a model from the target architecture.
}
\end{footnotesize}
\end{table}

{\em Purely static} approaches work at compilation time.
They might be applied by the compiler, either automatically, i.e., without user
intervention~\cite{Cong2012,Jain16,Luk09,Rossbach13,Poesia17,Tang2013}, or not.
In the latter case, developers can use annotations~\cite{Mendonca17}, domain
specific programming languages~\cite{Luk09,Rossbach13} or library
calls~\cite{Augonnet11} to indicate where each program part should run.
In Table~\ref{tab:relatedwork}, techniques implemented at either the
compiler or library levels are purely static.
{\em Purely dynamic} approaches take into account runtime information.
They can be implemented at the architecture
level~\cite{Rangan2009,Lukefahr2016,Joao12,Craeynest12,Yazdanbakhsh2015},
or at the virtual machine (VM)/OS level~\cite{Petrucci15,Nishtala17,Zhang2016,Gaspar2015,SomuMuthukaruppan2014,Barik16}.
By leveraging runtime information, the system can use environment information, 
unknown at compilation time, to solve \spha.
However, there may be some overhead on accurately collecting and processing
runtime data.
Besides, because scheduling decisions are taken on-the-fly, usually the
scheduler cannot spend much time weighting choices.
Thus, even though these algorithms use runtime information, they might still
take suboptimal decisions.
Approaches that mix static and dynamic techniques are called {\em hybrid}.
Astro is a hybrid method.
Other hybrid approaches to this problem
exist~\cite{Piccoli14,Cong2012,Tang2013}.
None of these previous work use any form of learning technique to
adapt the program to runtime conditions, as Table~\ref{tab:relatedwork}
indicates in the column {\em Learn}.
Once guards are created, they always behave on the same way.
That is the main difference between these previous approaches and the Astro method.

\section{Conclusion}
\label{sec:con}

This paper has presented Astro, a program scheduler for
big.LITTLE architectures.
Astro uses machine learning to adapt a program to runtime conditions.
However, it departs from previous approaches, also based on machine learning,
because it takes program characteristics into consideration.
Astro relies on the compiler to identify program regions that
contain similar syntactical features.
We classify these features in sets called program phases, and track, at
runtime, which program phase is currently valid.
When combined with dynamic data, this information lets a neural network
train the program, so to maximize some metric of efficiency, such as energy
or runtime.
By combining static and dynamic information, we are, effectively, building
architecture-aware code optimizations for parallel programs.

\bibliography{ref}


\end{document}